\documentclass[11pt]{article}
\usepackage{graphics}
\usepackage{epsfig}
\usepackage{epstopdf}
\usepackage{amsmath}
\usepackage{array}
\usepackage{setspace}
\begin{document}
\date{}

\title{{\bf Dynamical system analysis and thermal evolution of the causal dissipative model}}

\author{Jerin Mohan N D, Krishna P B, Athira Sasidharan and Titus K Mathew\\
Department of Physics, Cochin University of Science and Technology, \\
Kochi-22, India. \\
jerinmohandk@cusat.ac.in;krishnapb@cusat.ac.in,\\athirasnair91@cusat.ac.in,titus@cusat.ac.in}

\maketitle

\begin{abstract}
 The dynamical system behaviour and thermal evolution of a homogeneous and isotropic dissipative universe are analyzed.
 The dissipation is driven by the bulk viscosity $\xi = \alpha \rho^s $ and the evolution of bulk viscous pressure is
 described using the full causal Israel-Stewart theory. We find that for $s=1/2$ the model possesses a prior decelerated
 epoch which is unstable and a stable future accelerated epoch. From the thermodynamic analysis, we have verified that the local
 as well as the generalised second law of thermodynamics are satisfied throughout the evolution of the universe. We also show that the
 convexity condition $S''<0$ is satisfied at the end stage of the universe which implies an upper bound to the evolution of the
 entropy. For $s\neq1/2,$ the case $s<1/2$ is ruled out since it does not predict the conventional evolutionary stages
 of the universe. On the other hand, the case $s>1/2$ does imply a prior decelerated and a late de Sitter epochs, but both of
 them are unstable fixed points. The thermal evolution corresponding to the same case implies that GSL is satisfied at both
 the epochs but convexity condition is violated by both, so that entropy growth is unbounded. Hence for $s>1/2$ the model
 does not give a stable evolution of the universe. 
\end{abstract}

%\begin{twocolumn}

\section{Introduction}
Astronomical observations (\cite{Riess,Perlmutter,Bennett,Riess2,Tegmark,Seljak,Komatsu}) have shown that the current universe
is expanding at an accelerating rate.
The most successful model which explains this recent acceleration
is the $\Lambda$CDM, which assumes the
cosmological constant $\Lambda,$ with equation of state
$\omega_{\Lambda}=-1$ as the cosmic component responsible for the acceleration.
But due to the huge
difference between the predicted and observed values of the cosmological constant and also due to the
surprising coincidence between the present
densities of the dark matter and dark energy \cite{Sami1}, attention has been turned towards dynamical
dark energy models \cite{Wang,Caldwell,Bamba}. However the nature and composition of dark energy is still
a mystery and its
possible coupling with matter \cite{wang1} is also unknown. Modified gravity
theories \cite{Dvali,Freese} have been proposed as alternative solutions.
Another interesting approach is to invoke viscosity in the dark matter sector which can produce adequate negative
pressure to cause the
late acceleration \cite{Brevik1,Brevik2,Avelino1,Athira}. In the Weinberg formalism \cite{Misner,Weinberg,Weinberg1}
of the imperfect fluid, the bulk viscous fluid can act as a source in Einstein field equation. Very recently it has been
shown that the viscosity of the dark
matter can alleviate the discrepancy in the values of the cosmological parameters when one use the large scale structure
(LSS) and Planck data \cite{anand1} to constrain the parameters in the respective cosmological models.

Physically,
bulk viscosity can be generated whenever a system deviates from the local thermodynamic equilibrium \cite{Wilson}.
In cosmic evolution, the viscosity arises as an effective pressure to restore the system back into the thermal equilibrium
whenever the universe undergoes  fast expansion or contraction \cite{Okumura}.
The bulk viscosity thus generated can cause a negative pressure similar to the cosmological
constant or quintessence \cite{Mathews,Avelino2}. Even though this is a possible realistic picture for the generation of
bulk viscosity,
its origin in the expanding universe is still not clearly understood. Some
authors have shown that different cooling rates of components of the cosmic medium can produce bulk viscosity
\cite{Weinberg2,Schweizer,Udey,Zimdahl3}. Another proposal is that bulk viscosity of the cosmic fluid may be the result
of the particle number non-conserving interactions \cite{Murphy,Turok,Zimdahl4}.

The simple way of accounting the bulk viscosity in the expanding universe is through the Eckart theory \cite{Eckart}, which
gives a linear relationship between the bulk viscous pressure and the expansion rate of the universe. Since it is limited to the first
order deviation from the equilibrium, the Eckart theory suffers from serious short comings like the violation of causality
\cite{Coley2,Israel1} and the occurrence of unstable equilibrium
states \cite{Hiscock1}.
But it has been used by several authors to model the bulk viscosity in
explaining the late acceleration of the universe \cite{Brevik1,Fabris,Barrow1,Colistete,Avelino1,Avelino2,Athira,Athira2},
primarily due to its simplicity. Such cosmological models lead to reasonably good description of the background evolution of the universe,
but become problematic while considering the structure formation scenario.

A more general theory, consistent with the relativistic second order evolution of the bulk viscous
pressure, was suggested by Israel and Stewart \cite{Israel1,IsraelStewart,IsraelStewart2} and is free from the shortcomings
of the Eckart formalism. The inclusion of the dissipative second order terms ensure causality in the Israel-Stewart model and
it also accounts for the stability of the corresponding solutions.
In the limit of vanishing relaxation time, the Israel-Stewart theory reduces to the Eckart theory.
In some recent dissipative cosmological models \cite{Piattella}, a truncated version of the Israel-Stewart theory
has been used
in which one omits the divergence terms in the expression for the evolution of the bulk viscous pressure. Strictly
speaking such an approximation is valid only when cosmic fluid is very close to the equilibrium state.

It was noted in \cite{Padmanabhan,Prisco} that both causal and non-causal dissipative models in the context of early inflation
of the universe have some critical issues which makes role of viscosity in the early universe
rather unlikely. But in the
context of the late evolution of the universe the bulk viscous models are promising. Based on the Eckart approach, the
late acceleration can be explained without invoking to any fictitious dark energy component
\cite{Brevik1,Brevik2,Avelino1,Athira}.
A dynamical system analysis of the same model can predict the conventional evolution of the late universe if the bulk viscous
coefficient is a constant \cite{Athira2}.
The background evolution of the bulk viscous universe using the full Israel-Stewart theory has been analyzed in our
previous work \cite{EPJC1} where we have obtained
analytical solutions which explain the late acceleration of the universe with a transition redshift, $z_T\sim0.52,$
which shows the feasibility of describing a late accelerating universe. The current status of the viscous models are described
in the review \cite{Brevikrev1}.

In the present work, our aim is two fold. Firstly to perform a dynamical system analysis of dissipative model of the late universe and
secondly to study the thermodynamic evolution of the model based on the IS theory. In both analyses we choose the viscosity as $\xi \propto \rho^s,$ where
the parameter take values $s=1/2$ or $s \neq 1/2.$
The first method is aimed at finding the critical points of the autonomous differential equations
which are obtained from the Friedmann equations consistent with the conservation conditions. The sign and properties
of the eigenvalues corresponding to these critical points will then determine the asymptotic stability of the model.
Our analysis show that, there exists an unstable critical point corresponds to prior decelerated universe and an asymptotically
stable critical point corresponding to a future accelerating epoch for $s=1/2.$
We also explore the status of the energy conditions, both the strong and dominant energy conditions, to check the physical feasibility of
the solutions corresponding to the respective
critical points. Further, we analyses the thermal evolution of the model where we check the status of the
generalized second law (GSL) and the convexity condition, $S^{\prime\prime} <0,$ where $S$ is the entropy and the prime
denotes a derivative with respect to a suitable cosmic variable.
In this context we found that
the end stage in this model is thermodynamically stable with an upper bound for entropy when $s=1/2,$ which indicates that our
universe behaves like an ordinary macroscopic system \cite{Pavon1}. 
Authors in reference \cite{Cruz1} have analyses the viscous model following Israel-Stewart approach, by considering an ansatz for the Hubble 
parameter and with varying barotropic equation of state and have shown, in contrary, that the end stage violates the convexity condition.
However for $s\neq 1/2$ the results, in the present model, are
not in favour of the evolution towards stable epoch of the universe.

The paper is organized as follows. In section (\ref{sec:1}), the Hubble parameter from the full causal Israel-Stewart
theory is obtained.
The dynamical behaviour of the bulk viscous model for $s=1/2$ and $s\neq1/2$ are studied in section (\ref{sec:2}).
The section (\ref{sec:3}) deals with the analysis of the thermodynamic
conditions during the evolution of the present model of the universe and our conclusions are given in section (\ref{sec:4}). The possibilities of attaining a pure de Sitter phase for $s=1/2$ is discussed in the Appendix.

\section{The causal viscous model}
\label{sec:1}
We consider a flat FLRW universe with viscous matter as the cosmic component. The basic equations governing the evolution of the universe are,
\begin{equation} \label{eqn:F1}
3H^{2}={\rho_m},
\end{equation}
\begin{equation}\label{eqn:Hdot1}
\dot H = -H^2 - \frac{1}{6} \left(\rho_m+3P_{eff} \right),
\end{equation}
where $H=\frac{\dot{a}}{a}$ is the Hubble parameter with $a$ is the scale factor, $ \rho_{m} $ is the matter density
and
\begin{equation}
\label{eqn:effectivep}
P_{eff}=p+\Pi,
\end{equation}
is the effective
pressure, $ p=(\gamma-1)\rho $ is the normal kinetic pressure with $ \gamma $ as the barotropic index and
$ \Pi $ is the bulk viscous pressure. The evolution of the density of the viscous fluid satisfies the
conservation equation,
\begin{equation} \label{eqn:con1}
\dot{\rho}_{m}+3H(\rho_{m}+P_{eff})=0.
\end{equation}
In the full causal IS theory, the evolution of the viscous pressure is given by,
\begin{equation}\label{eqn:IS1}
\tau\dot\Pi+\Pi=-3\xi H-\frac{1}{2}\tau\Pi\left(3H+\frac{\dot\tau}{\tau}-\frac{\dot\xi}{\xi}-
\frac{\dot{T}}{T}\right),
\end{equation}
where $\tau$, $ \xi $ and $ T $ are the relaxation time, bulk viscosity and temperature respectively and are generally functions of the density of
the fluid, defined by the following equations \cite{Maartens},
\begin{equation}
\label{eqn:tau}
\tau=\alpha\rho^{s-1}, \, \, \,
\xi=\alpha\rho^{s}, \, \, \,
T=\beta\rho^{r},
\end{equation}
Here $\alpha$, $\beta$ and $s$ are all positive constant parameters
and $ r=\frac{\gamma-1}{\gamma} $. For $\tau=0$, the differential equation (\ref{eqn:IS1}) reduces to the
simple Eckart equation,  $\Pi=-3\xi H.$
Friedmann equation (\ref{eqn:F1}) can be combined with 
(\ref{eqn:con1}) and (\ref{eqn:effectivep}) to express the bulk viscous pressure $\Pi$ as,
\begin{equation}
\label{eqn:pi}
\Pi=-\left[2\dot{H}+3H^2+(\gamma-1)\rho\right].
\end{equation}
Following this, the bulk viscosity evolution in (\ref{eqn:IS1}) can be expressed as,
\begin{eqnarray}
\ddot H + \frac{3}{2}\left[1+(1-\gamma)\right] H \dot H + 3^{1-s} \alpha^{-1} H^{2-2s} \dot H \nonumber\\
- (1+r)H^{-1} {\dot H}^2 + \frac{9}{4}(\gamma -2)H^3+ 
\frac{1}{2}3^{2-s}\alpha^{-1}\gamma  H^{4-2s} =0.
\end{eqnarray}
For
$\gamma=1$ corresponding to non-relativistic matter and taking $s=\frac{1}{2}$ \cite{ChimentoJacubi}, the above equation
admits solution \cite{EPJC1} of the form,
\begin{equation}
\label{eqn:Hubbleparameter}
H=H_0\left(C_1 a^{-m_1}+C_2a^{-m_2}\right),
\end{equation}
where $ H_0$ is the present Hubble parameter and the other constants are \cite{EPJC1},
\begin{equation}
\label{consatntC1}
C_{1;2}=\frac{\pm 1+\sqrt{1+6\alpha^{2}} \mp \sqrt{3}\alpha\tilde{\Pi}_0}{2\sqrt{1+6\alpha^{2}}},
\end{equation}
\begin{equation}
\label{constantm1}
m_{1;2=}\frac{\sqrt{3}}{2\alpha}\left(\sqrt{3}\alpha+1 \mp \sqrt{1+6\alpha^{2}}\right).
\end{equation}
Here $ \tilde{\Pi}_0=\frac{\Pi_0}{3H_{0}^{2}} $ is the dimensionless
bulk viscous pressure parameter, with $\Pi_0$ as the present value of $\Pi.$ The model parameters up to 1$\sigma$ level were
estimated by contrasting the model with the supernovae data \cite{EPJC1} and are given in
table \ref{Table:ParametersFIS}. We find $m_1=0.31,$ $ m_2=5.29.$
Since $m_1<1 \, \textrm{and} \, m_2>1,$ the expansion rate will be dominated by $a^{-m_2}$ in the early epoch, while the
term $a^{-m_1}$ dominates in the late epoch. Hence in the limit $a\rightarrow 0,$ the
deceleration parameter $q,$ becomes $q=-1-\dot H/H^2 \to -1+m_2 > 0,$
which implies a prior decelerated expansion phase. But in the limit
$a\rightarrow\infty,$ it turn out that
$q\to -1+m_1<0,$ which implies a late
accelerating phase of expansion and therefore the model predicts a transition into the late accelerating epoch.
However, since $m_1$ is a positive quantity, the deceleration parameter will general be greater than $-1,$ but owing to the smallness 
of $m_1$ it can approach a value near to $-1$ corresponding to a pure de Sitter epoch \cite{EPJC1}. We will look into this point at later section.
Further since the model assumes a single cosmic component, it follows that $\Omega_{total} \sim \Omega_{darkmatter}.$ From (9) the matter density parameter $\Omega_m$ is obtained as 
\begin{equation}
\Omega_m=\frac{\rho_m}{\rho_{critical}}=\frac{ H^2}{H_0^2}=(C_1 a^{-m_1}+C_2 a^{-m_2})^2.
\end{equation}
The matter density parameter in the present time $\Omega_{m0},$ corresponding to $a=1$ and is,
\begin{equation} 
\Omega_{m0}=(C_1+C_2)^2=1.
\end{equation}
\begin{table}	
	\center{\begin{tabular}{@{}lccccc@{}}\mbox{} \\\hline\hline\\                         
			$ H_{0}$ & $ \alpha $  & $ \tilde{\Pi}_{0}$ &
			$ \chi^{2}_{min} $ & $ \chi^{2}_{d.o.f.} $\\ \hline \vspace{0.05in}
			$70.29$  & $0.665^{+0.030}_{-0.025}$ & $-0.726^{+0.01}_{-0.01}$  & $310.29$ & $1.020$ \\\hline\hline
	\end{tabular}}\\
	\caption{The best estimated values of the model parameters and the $\chi^2$ minimum value in the bulk viscous matter dominated universe using the
		full IS theory as per the earlier work \cite{EPJC1}. We have used the Supernovae data.} 
	\label{Table:ParametersFIS}
\end{table}

\section{Dynamical system analysis}
\label{sec:2}
We will now consider the dynamical system analysis \cite{Ellis} of the model. For this we define the
following dimensionless variables,
\begin{equation}
	\label{eqn:dimensionlessdensityphasespace}
	\Omega=\frac{\rho_m}{3H^2}, \, \, \,
	\tilde{\Pi}=\frac{\Pi}{3H^2}, \, \, \, \textrm{and} \, \, \,
	H(t)dt=d\tilde{\tau},
\end{equation}
where the last relation is equivalent to a new time variable.
The (\ref{eqn:Hdot1}), (\ref{eqn:con1}) and the IS equation (\ref{eqn:IS1}) can then be re-written as,
\begin{equation}
	\label{eqn:H'}
	H'=-H\left[1+\frac{1}{2}(\Omega+3\tilde{\Pi})\right],
\end{equation}
\begin{equation}\label{eqn:omega'}
	%\label{eq:omega'}
	\Omega'=(\Omega-1)(\Omega+3\tilde{\Pi}),
\end{equation}
and
\begin{equation}
	\label{eqn:Pi'}
	\tilde{\Pi'}=-3\Omega- \tilde{\Pi}\left[\frac{3}{2}\left(2+\frac{\tilde{\Pi}}{\Omega}\right) +
	\frac{H^{1-2s}}{\alpha (3\Omega)^{s-1}}-\Omega-3\tilde{\Pi}-2\right],
\end{equation}
where the $'prime'$ denotes a derivative with respect to the new variable $\tilde{\tau}.$ Since $H$ is always positive for an
expanding flat universe, the above
equations are well defined. The above three dynamical equations constitute the
evolution of the system in a phase space
described by the
variables $(H,\Omega, \Pi).$ We are considering a universe with single component, the viscous matter, implying that $\Omega=1.$ Then the phase space becomes two dimensional
with variables $(H, \tilde{\Pi}).$
The critical parameter in studying the evolution is $s.$ We have found exact solutions for $s=1/2$ in the previous section.
However for analyzing dynamical system behaviour
we will consider choices $s\neq1/2 $ also in accounting for the bulk viscosity.
\subsection{Choice 1. $s=1/2$}
For this choice
(\ref{eqn:H'}) and (\ref{eqn:Pi'}) decouple from each other and as a result the phase space will effectively
reduces to one dimension and
(\ref{eqn:Pi'}) represents the evolution of this single dimensional phase space. More over in the present case,
since $\Omega=1,$ (\ref{eqn:Pi'})
can be expressed in a much simpler form in terms of the equation of state, $\omega=\tilde{\Pi}/\Omega$ as,
\begin{equation}\label{eqn:autoeq1}
	\omega'=\frac{3}{2}(\omega-\omega^+)(\omega-\omega^-),
\end{equation}
where \begin{equation}
	\label{eqn:omegapm}
	\omega^{\pm}=\frac{1}{\sqrt{3}\alpha}[1\pm\sqrt{1+6\alpha^2}]
\end{equation}
are the fixed points. The equation (\ref{eqn:omegapm}) implies that, $\omega^+>0$ and $\omega^-<0$ for all $\alpha>0.$ The early
phase corresponding to $\omega^+$
is decelerating. If $\alpha$ is sufficiently large then $\omega^-< -1/3$ and consequently the late epoch of the universe will be accelerating.
So depending on the value of the parameter $\alpha$ the equation of state can assume values accordingly. For a range 
$\frac{2\sqrt{3}}{17} \leq \alpha \leq \frac{2}{\sqrt{3}}$ the equation of state vary between to $-1/3 > \omega \geq -1.$ So an asymptotic de Sitter
epoch ($\omega \to -1$) is possible only if $\alpha$ assumes the upper limit value around $\frac{2}{\sqrt{3}}.$
For the best estimated value of the model parameter, we have obtained that
$\omega^+=2.52$ and $\omega^-=-0.79$ and are corresponding to a prior decelerated phase in which the viscous matter assumes a stiff fluid 
nature and a late accelerated epoch, in which the matter assumes a quintessence nature respectively. So for the case with 
$\gamma=1$ (the barotropic index) and $\epsilon=1$ ($\gamma$ and $\epsilon$ appears in the general equation of relaxation time (\ref{eqn:relaxationtimegeneral}) given in the Appendix) the late universe with bulk viscous matter can be accelerating but it will not approach a pure de 
Sitter epoch like the standard $\Lambda$CDM.

The deceleration parameter corresponding to the equilibrium points can be obtained using the relation
$1+q=\frac{3}{2}(1+\omega),$ through which we arrive at
$q^+\sim4.28$ and
$q^-\sim-0.69.$ Taking account of these facts, it is possible to re-write the general solution given in (\ref{eqn:Hubbleparameter}) as,
\begin{equation}
	\label{eqn:Hubbleparameterwithq}
	H=H_0\left(C_1 a^{-(1+q^-)} + C_2 a^{-(1+q^+)} \right).
\end{equation}
Using this the transition from the decelerated to the current accelerated phase of expansion can easily be explained. 
The transition redshift $z_T.$ can be obtained 
using (\ref{eqn:Hubbleparameterwithq}) as,
\begin{equation}
	\label{Transition redshift}
	z_{T}=\left(-\frac{C_1 q^+}{C_2 q^-}\right)^{-\frac{1}{q^+ - q^-}}\sim 0.52,
\end{equation}
where the numerical value is corresponding to the best estimated values of the model parameters and is found to be in the 
WMAP range $z_T=(0.45-0.73)$ \cite{U.Alam}.

Without knowing the analytical solution, it is possible to analyze the cosmic evolution from (\ref{eqn:autoeq1})
in a transparent way
by drawing the phase diagram of $\omega,$ namely plotting $\omega^{\prime}$ versus $\omega.$ Since the phase space is one dimensional, we interpret 
(\ref{eqn:autoeq1})
as a vectorfield on a single line \cite{Awad1}. The evolution of $\omega$ is represented by the direction of the change of $\omega$ along the axis and is determined
by the sign of $\omega^{\prime}.$
A small variation in $\omega$ is expressed
as $\delta\omega=\omega'\delta\tilde{\tau},$ so that for $\omega$ flows towards the increasing direction of $\tilde{\tau}$ (right) if $\omega'>0$ and
flows towards the decreasing direction (left) if $\omega'<0.$ Perturbations in the $\omega$ space around the critical point, $\omega_c \, \, (\omega^+  \, \textrm{or}  \,
\omega^-)$ propagates with a rate \\
\begin{equation}
	\frac{d}{d\tilde{\tau}}\left(\delta\omega\right)=\omega'=f(\omega)=f(\omega_c+\delta \omega).
\end{equation}
The Taylor series expansion around $\omega_c$ can be written as,
$f(\omega_c+\delta \omega)=f(\omega_c)+\delta \omega f'(\omega_c)+O(\delta \omega^2), $ where
$f'(\omega_c)=\frac{d}{d\omega}f(\omega)|_{\omega_c},$ from which we
get $\frac{d}{d\omega} (\delta \omega)=\delta \omega f'(\omega_c).$ By linearising $\delta \omega$ about the critical point $\omega_c$ we get,
\begin{equation}
	\label{eqn:stability1D}
	\delta \omega(\tilde{\tau}) \propto e^{f'(\omega)\tilde{\tau}}.
\end{equation}
The above equation tells us that the stability of critical points is determined by the slope, $f^{\prime}(\omega).$ If $f'(\omega_c)>0,$ then any small
disturbance around the critical point grow exponentially and hence it becomes unstable (repeller). On the other hand, if $f'(\omega_c)<0,$
all small disturbances around critical point decay exponentially and it will be a stable one(attractor). The critical point will be semi
stable, if the slope $f'(\omega_c)$ changes its sign at the critical point.

\begin{figure}
	\centering
	\includegraphics{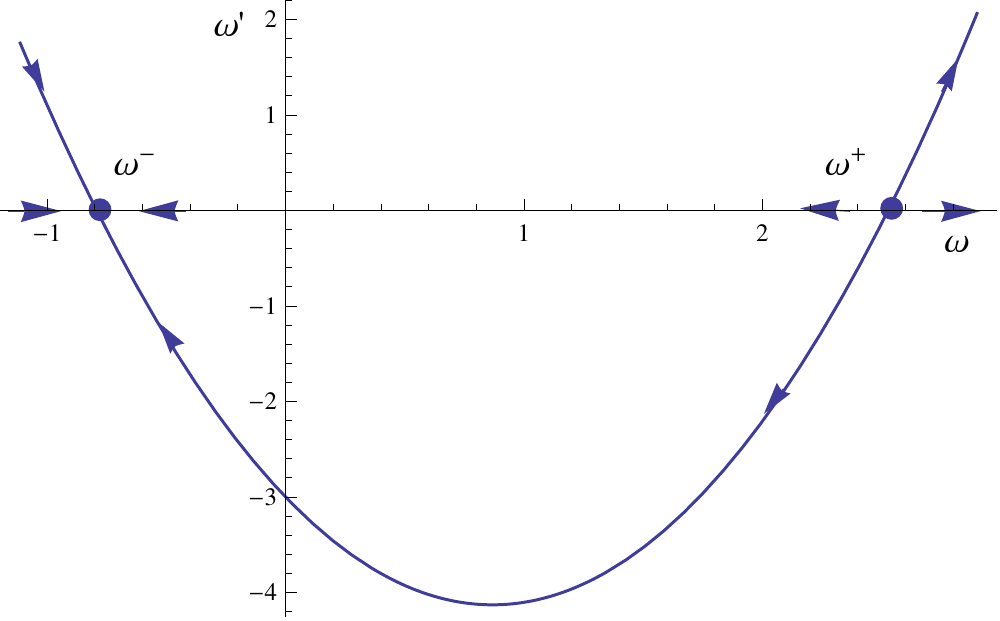}
	\caption{The one dimensional phase portrait of evolution of $\omega'$ versus $\omega$ in the bulk viscous matter dominated universe using the full causal IS theory for
		best estimated value of the model parameter, when $s=1/2.$}
	\label{plot:onedimensionalplot}
\end{figure}

The slope corresponding to (\ref{eqn:autoeq1}) can be obtained as,
\begin{equation}
	\label{eqn:omega''1D}
	f'(\omega)=\frac{3}{2}\left[2\omega-\omega^+ - \omega^-\right].
\end{equation}
For the best estimated values of the model parameters, it is clear that the condition, $\omega^- < \omega < \omega^+$ is always be satisfied.
Then, at the critical points $\omega^{\pm}$ the slope will satisfy the conditions,
\begin{equation}
	\label{eqn:omega''+}
	f'(\omega^+)=\frac{3}{2}\left[\omega^+ - \omega^-\right]>0 \,\, for  \,\,\alpha>0,
\end{equation}
\begin{equation}
	\label{eqn:omega''-}
	f'(\omega^-)=\frac{3}{2}\left[\omega^- - \omega^+\right]<0  \,\,for  \,\,\alpha>0,
\end{equation}
which indicates that $\omega^+$ is an unstable fixed point while $\omega^-$ is a stable fixed point. Hence the universe will evolves from an unstable decelerated epoch to the
stable accelerated epoch. So in effect we get a qualitative description of the behaviour of the cosmological evolution without
relying on the exact solution. The phase
portrait is shown in figure \ref{plot:onedimensionalplot}.

The exact solutions corresponding to the fixed points $\omega^\pm$ follows from  (\ref{eqn:H'}) are,
\begin{equation}
	\label{eqn:H1D}
	H_{\omega^\pm}=\frac{1}{(1+q^{\pm})t}, \quad
	a_{\omega^\pm}=a_0 t^{\frac{1}{(1+q^\pm)}}.
\end{equation}
For $\omega^+ $ we have $\frac{1}{(1+q^{+})}<1,$ indicating a decelerating solution, while for $\omega^-$ we have
$\frac{1}{(1+q^{-})}>1$ implying an accelerating solution.
The density and pressure then follows the evolution,
\begin{equation}
	\label{eqn:densityandpressure1D}
	\rho_{\omega^\pm}=\frac{3}{(1+q^\pm)^2 t^2}, \,\,\,\,\, \tilde{\Pi}_{\omega^\pm}=\frac{3\omega^\pm}{(1+q^\pm)^2 t^2}.
\end{equation}
For $\omega^+$ the pressure $\tilde\Pi >0$ and for $\omega^-$ it becomes negative, $\tilde\Pi <0,$ implying the
generation of negative pressure in the late acceleration epoch. \\

It is essential to know the status of the energy conditions \cite{Visser} which characterize the feasibility of different solutions.
The strong energy condition (SEC) implies that
$\rho+3P_{eff}\geq0$. The violation of SEC indicates an accelerating expansion of the universe.
The dominant energy condition (DEC) implies that $\rho+P_{eff}\geq0.$ A violation of DEC causes the breakdown of the generalised second law of thermodynamics in normal
case. However, if there occur dissipative effects in the cosmic fluid, the GSL can still be satisfied even when DEC is violated \cite{Pavon12}.
In term of equation of
state SEC and DEC can translated as, $1+3\,\omega\geq0$ and $1+\omega\geq0$ respectively. For the best estimated values of the parameters, it can easily be seen that
both SEC and DEC are satisfied by $\omega^+.$ In the case of the fixed point $\omega^-,$ SEC is violated, since it represents an accelerating solution, but DEC is satisfied
as it is a physically feasible epoch. All these facts are summarized in table \ref{Table:FIS1/2}. \\
\begin{table}	
	\center{\begin{tabular}{@{}lccc@{}}\mbox{} \\\hline\hline\\                           
			Critical points $\rightarrow$&$\omega^+$ & $\omega^-$ \\\hline
			$\omega$  & $2.52$& $-0.79$ \\
			$q$& $4.28$ & $-0.69$  \\
			Stability & Unstable  & Stable \\
			SEC & Yes & No \\
			DEC & Yes & Yes   \\\hline\hline
	\end{tabular}}\\
	\caption{Qualitative properties of the critical points $\omega^+$ and $\omega^-$ in the dynamic system of the bulk viscous model using the full causal Israel-Stewart theory for the best estimated values of the model parameters, when $s=1/2$.} 
	\label{Table:FIS1/2}
\end{table}

A similar case of the non-validity of the strong energy condition for a future accelerating epoch was also pointed out by
Barrow \cite{Barrow2} using the Eckart formalism to account for the viscosity. 
\subsection{Choice 2. $s\neq1/2$} 
Unlike in the case of $s=1/2$ a complete description of dynamic evolution is difficult for $s\neq1/2.$ However, we can get
a qualitative description of the evolution by extracting the information from the equilibrium points.
In this case we have a two dimensional space $(H,\Pi).$ For simplicity we define a variable.
\begin{equation}
	\label{eqn:h}
	h=H^{1-2s}.
\end{equation}
The dynamical equations (\ref{eqn:H'}) and (\ref{eqn:Pi'}) then become,
\begin{equation}
	\label{eqn:h'}
	h'=-\frac{3}{2}(1-2s)(1+\omega)h,
\end{equation}
\begin{equation}
	\label{eqn:omega'2D}
	\omega'=-3\left[1+\omega\left(\frac{3^{(-s)}}{\alpha}h-\frac{\omega}{2}\right)\right].
\end{equation}
having two critical points,
\begin{equation}
	\label{eqn:P2}
	P_1:\quad h=0,\quad\qquad \quad \omega=\sqrt{2},
\end{equation}
\begin{equation}
	\label{eqn:P3}
	P_2:\quad h=\frac{3^s\alpha}{2},  \qquad \quad  \omega=-1.
\end{equation}

For $s<1/2$ the critical point $P_1$ with $h=0$ implies a static universe with the Hubble parameter $H=0.$
The critical point $P_2$  corresponds to a de Sitter epoch at which the Hubble parameter is a non-zero constant. Since
the first phase $P_1$ is a static one it will not imply any further evolution. Hence the case $s<1/2$
fails to predict a prior decelerated epoch, it is not worth exploring any further.
On the other hand, for $s>1/2,$ the fixed point $P_1$ is representing a prior decelerated epoch with infinitely large
Hubble parameter and $P_2$  corresponds to a late de Sitter epoch. However the equation of state corresponding to the
prior decelerated epoch is greater than one, implying that the matter is of stiff nature at this epoch.
We will restrict to the case $s>1/2$ in our further analysis.
Regarding the energy conditions, it is found that both SEC and DEC are satisfied at $P_1,$ and hence it corresponds
to physically feasible decelerating epoch.
The fixed point $P_2,$ satisfies DEC but violates SEC as it is corresponding to an accelerating epoch. \\

To determine the stability property of the critical points, we first linearize (\ref{eqn:h'})
and (\ref{eqn:omega'2D})
about the critical points and obtain the Jacobian matrix as,
\begin{equation}
	J(h, \omega)=\left[ \begin{array}{cc}
		-\frac{3}{2}(1-2s)h & -\frac{3}{2}(1-2s)(1+\omega) \\ 
		-3\left(\frac{3^{-s} h}{\alpha}-\omega\right) & -\frac{3^{1-s}\omega}{\alpha}
	\end{array}  \right]
\end{equation}

Diagonalising the Jacobian matrix, we obtain the eigenvalues
\begin{equation}
	\label{eqn:eigenvalueP2}
	\lambda_1^{\pm}=\frac{3^{1-s}}{\sqrt{2}\alpha}\left[-1\pm \sqrt{1+9^s (\sqrt{2}+2)(2s-1)\alpha^2}\right],
\end{equation}
\begin{equation}
	\label{eqn:eigenvalueP3}
	\lambda_2^{+}=\frac{3^{1-s}}{\alpha}, \qquad \lambda_2^{-}=\frac{3^{1+s}}{4}(2s-1)\alpha,
\end{equation}
for $P_1$ and $P_2$ respectively.
Here we restrict $\alpha$ to the range $0<\alpha<1.$
The fixed point $P_1$ is a saddle one,
since the eigenvalues are, $\lambda_1^+>0,$ $\lambda_1^-<0,$ while $P_2$ is found to be unstable, since its eigenvalues are both positive,
$\lambda_2^+>0,$ $\lambda_2^->0.$ The saddle nature of the early decelerated phase implies that the system will continue the evolution further.
For the sake of completeness, it may be noted that, for $s<1/2,$ the fixed point $P_1,$ which corresponds to a static universe, is found to be stable
since $\lambda_1^+<0$ and $\lambda_1^-<0$ and $P_2$ is a saddle point as the eigenvalues satisfies, $\lambda_2^+>0$ and $\lambda_2^-<0.$
All these facts are summarised in table \ref{Table:FIS3/21/4}. 

The fixed point $P_2$  corresponds to a solution given by,
\begin{equation}
	a=a_0 e^{\bar{H}_0 t},
\end{equation}
where $\bar{H}_0=\left(\frac{3^s \alpha}{2}\right)^{\frac{1}{1-2s}}$ with $s>1/2.$ This is a de Sitter type solution, ensuring accelerated expansion.
It is not possible to get any corresponding exact solution for $P_1$ as the Hubble parameter in this case is infinity.
Even though an exact solution for $P_1$ is impossible, an approximate solution can be obtained. For this, first express the 
(\ref{eqn:h'}) in terms of $H$ and then through a simple integration we arrive at,
\begin{equation}
	\label{eqn:HP1P2}
	H \sim e^{-(1+q) \tau},
\end{equation}
from which it  is  evident that as $\tau\rightarrow-\infty,$  $H\rightarrow\infty.$
Integrating the above equation by changing the variable
from $\tau$ to $t$ using (\ref{eqn:dimensionlessdensityphasespace}), we get the scale factor as $a\sim t^{\frac{1}{(1+q)}}$ and the corresponding pressures is
$\tilde{\Pi}\sim \frac{3\,\omega}{(1+q)^2 t^2}$.

\begin{table}	
	\center{\begin{tabular}{@{}lccc@{}}\mbox{} \\\hline\hline\\                              
			Critical points $\rightarrow$& $P_1$ &$P_2$ \vspace{0.1cm}  \\\hline\\
			\vspace{0.13cm}
			$q$& $2.62$ &$-1$   \\
			$s>1/2$  &  Saddle  & Unstable   \\
			$s<1/2$ & Stable & Saddle \\
			SEC & Yes & No  \\
			DEC & Yes & Yes \\\hline\hline
		\end{tabular}}\\
			\caption{Qualitative properties of the critical points $P_1$ and $P_1$ in the dynamic system of the bulk viscous model using the full causal Israel-Stewart theory, when $s\neq1/2.$} 
			\label{Table:FIS3/21/4}
		\end{table}
\section{Thermodynamic analysis}
\label{sec:3}
This section is devoted to the analysis of the evolution of entropy. Viscosity can cause entropy generation and the local
entropy thus generated can be obtained as \cite{Weinberg},
\begin{equation}
\label{eqn:local entropy}
T\nabla_{\nu}S^{\nu}=\xi(\nabla_{\nu}u^{\nu})=9H^{2}\xi,
\end{equation}
where $ T $ is the temperature and $ \nabla_{\nu}S^{\nu} $ is the rate of generation of entropy in unit volume.
According to second law of thermodynamics, the entropy must always increase, i.e.
$T\nabla_{\nu}S^{\nu}\geq0,$
which implies
that $ \xi\geq 0.$ For $s=1/2,$ from (\ref{eqn:tau}) and
(\ref{eqn:Hubbleparameter}), it follows
$\xi=\sqrt{3} \alpha H.$ Since both, $\alpha$ and $H$ are always positive definite in the present case
the local second law will be satisfied. Then it is easy to conclude that the local second law will be satisfied at the
critical points $\omega^+$ and $\omega^-$ since they
are the critical points corresponding to the case $s=1/2.$ As there are no analytical solutions for $s\neq1/2,$ it is impossible to make a similar analysis.

Now we turn to the more general aspects of the entropy evolution, namely the status of the generalised second law (GSL) and the behaviour of the second order
derivative of entropy.
An ordinary macroscopic system evolving towards a state of stable
thermodynamic equilibrium must satisfy the conditions,
\begin{equation}\label{eqn:ddots}
S^{\prime} \geq 0, \hspace{0.24in} \textrm{and} \hspace{0.24in} S^{\prime\prime}<0, \, \, \textrm{at least in the long run}
\end{equation}
where $'prime'$ denotes a derivative with respect to suitable cosmological variable like cosmic time or scale factor.
The first condition refers to the GSL and the second one is the convexity condition implying
an upper bound to the growth of entropy. In reference \cite{Pavon1}, the authors have shown that our universe seems to
behave like an ordinary macroscopic system which obeys the above conditions. The consideration of the entropy evolution in the standard $\Lambda$CDM model
also supports this \cite{Krishna}.

According to GSL, the total entropy must always increase, i.e.,
\begin{equation}
S^{\prime}=S^{\prime}_m + S^{\prime}_h \geq 0,
\end{equation}
where $S_m$ and $S_h$ are the matter entropy and horizon entropy respectively and the $'prime'$ denotes the derivative with respect to
scale factor. The entropy of the Hubble horizon is defined as \cite{Davis},
\begin{equation}
\label{eqn:S_h}
S_{h}=\frac{A}{4l_{p}^2}k_B=\frac{\pi c^2}{{l_{p}^2}H^2}k_B,
\end{equation}
where $A=4\pi c^2/H^2$ is the area of the Hubble horizon of a spatially flat FLRW universe, $k_B$ is the Boltzmann constant,
$l_p$ is the Planck length and $c$ is
the velocity of light.
We have the derivative of the horizon entropy with respect to the scale factor as,
\begin{equation}
\label{eqn:S'_h}
{S'_{h}}=\frac{-2{\pi}c^2{H'}}{{l_{p}^2}H^3}k_B.
\end{equation}
The variation in the entropy of matter, $S'_m$ can be obtained from the Gibb's relation,
\begin{equation}
T_{m}S'_{m}=E'+P_{eff}V',
\end{equation}
where $T_m$ is the temperature of the viscous matter, $E=\rho_m V$ is its total energy and $V=\frac{4\pi c^3}{3H^3}$ is
the volume
enclosed by the Hubble horizon. Using the
Friedmann equation and assuming thermal equilibrium so that $T_m=T_h,$ where $T_{h}=\frac{H\hbar}{2\pi}{k_B}$ the Hawking temperature of the horizon,
we get
\begin{equation}
\label{eqn:S'_m}
S'_m=-\frac{c^5H'}{GH^2}\frac{q}{T_h}.
\end{equation}
Adding (\ref{eqn:S'_h}) and (\ref{eqn:S'_m}),
we get the rate of change of total entropy as,
\begin{equation}
\label{eqn:S'_general}
%S'=\frac{-2{\pi}c^2{H'}}{{l_{p}^2}H^3}\left( q +\frac{1}{\kappa} \right),
S'=\frac{-2{\pi}c^2{H'}}{{l_{p}^2}H^3}\left( q + 1 \right).
\end{equation}
\begin{figure}
	\centering
	\includegraphics{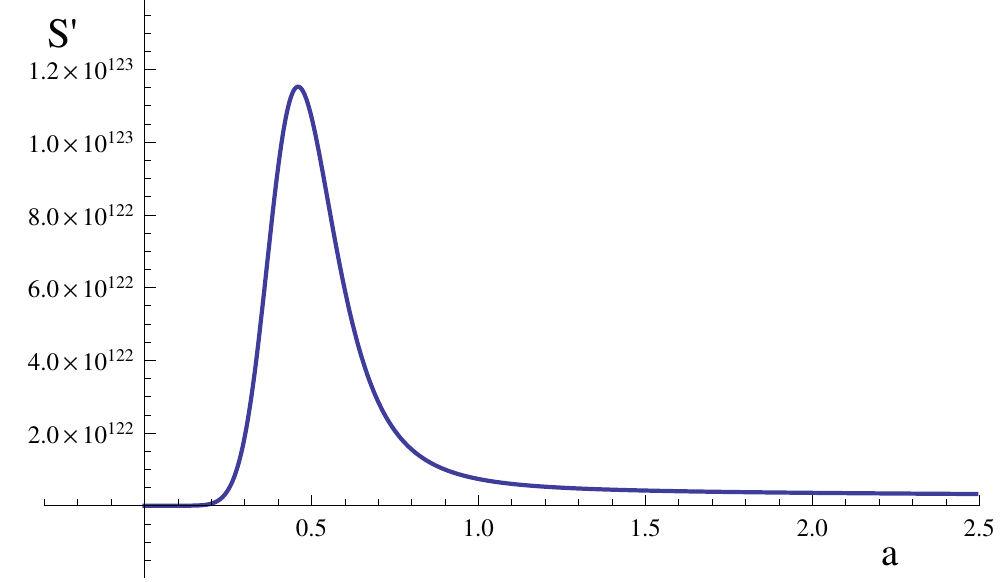}
	\caption{The evolution of $S' $ in units of $k_B$ with scale factor $a$ in the bulk viscous matter dominated universe using the full causal
		IS theory for best estimated values of the model parameters, when $s=1/2$.}
	\label{plot:S'_general}
\end{figure}
For $s=1/2,$ it is evident from general solution (\ref{eqn:Hubbleparameter}) that $H^{\prime}<0$ and also $(1+q)>0$ and
hence the GSL is valid. The evolution of $S^{\prime}$ is shown in figure \ref{plot:S'_general}
and is such that the first increase  occurs during the decelerated
epoch and then it decreases during the accelerated epoch. The figure 2 shows that the slope of the curve changes drastically around the transition redshift. The maximum of $S^{\prime}$  corresponds to the transition from
deceleration to acceleration epoch. It is then quite natural to expect that GSL will be satisfied at the corresponding
critical points, $\omega^+$ and $\omega^-.$ The Hubble parameter corresponding to these
fixed points is $H_{\omega^\pm}=\frac{1}{(1+q^\pm)a^{1+{q^\pm}}},$ implying that $H_{\omega^\pm}^{\prime}<0$ and hence
GSL is valid at both the points as expected.\\

For finding the status of GSL for $s>1/2,$ (we restrict to this case, since as noted earlier $H=0$ for the critical point corresponding to $s<1/2$)
we change the variable from scale factor to newly defined time,
$\tilde \tau.$ Following (\ref{eqn:HP1P2}) satisfied by $P_1$ we can rewrite
the entropy derivative
in (\ref{eqn:S'_general}) as,
\begin{equation}
\label{eqn:S'_P1P2}
\frac{dS}{d\tilde{\tau}}=\frac{2{\pi}c^2}{{l_{p}^2}}\left(1+ q\right)^2e^{2(1+q)\tilde{\tau}},
\end{equation}
and is always greater than zero. Hence
GSL is satisfied at $P_1.$
The validity of GSL at $P_{2}$ is straight forward since it represents a de Sitter epoch at which the Hubble parameter is
a constant implying  $S^{\prime}=0.$

Now will check the status of the convexity condition of entropy,
$S^{\prime\prime}<0,$ in this model.
This condition should be satisfied at least in the final stage of the evolution for the maximisation of entropy \cite{Pavon1}.
Taking the derivative of
$S^{\prime}$ in (\ref{eqn:S'_general}) with respect to the scale factor, we get
\begin{equation}
\label{eqn:S''_general}
S''=\frac{-2\pi c^2}{{l_p}^2}\left[\frac{H'}{H^3}q'+\left(q+1\right)\left( \frac{H''}{H'}-\frac{3H'}{H}\right)\right].
\end{equation}
\begin{figure}
	\centering
	\includegraphics{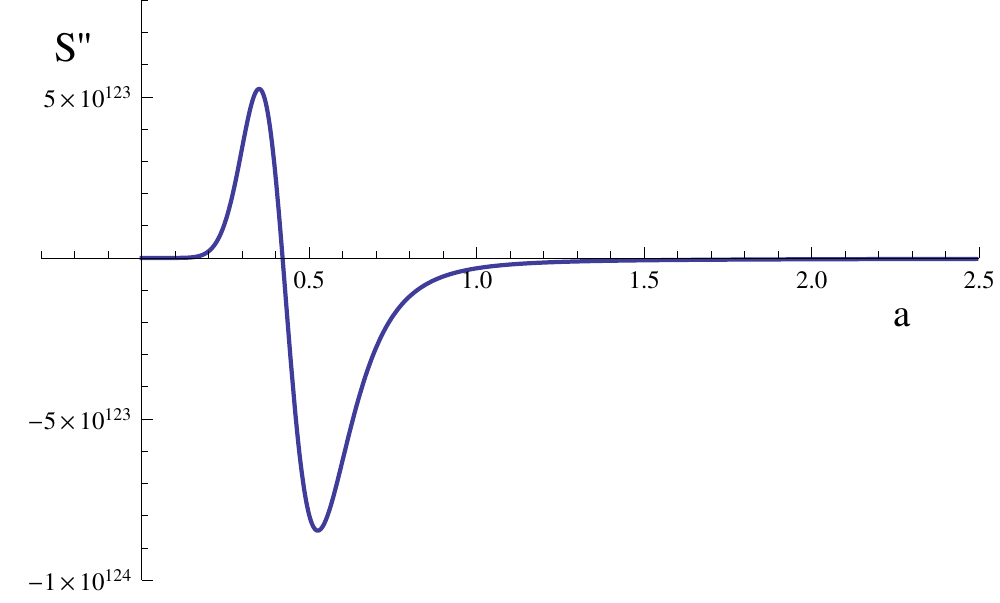}
	\caption{The evolution of $S'' $ in units of $k_B$ with scale factor $a$ in the bulk viscous matter dominated universe using the full causal IS theory for estimated values of the model parameters, when $s=1/2.$}
	\label{plot:S''_general}
\end{figure}
For $s=1/2$ the evolution of $S^{\prime\prime}$ can be obtained by substituting the Hubble parameter from (\ref{eqn:Hubbleparameter}).
The net result is plotted in
figure \ref{plot:S''_general}.
It shows that $S''>0$ during the early phase of evolution, while $S^{\prime\prime}<0$ in the later epoch and asymptotically approaches zero from below. The $S''$ changes its sign around the transition period.
Hence the convexity condition is fulfilled in the long run of the expansion of the universe.
This indicates the maximisation of entropy of the universe and hence
entropy is bounded. The boundedness of the entropy rules out the presence of any instabilities at the end stage \cite{Callen1}.
The behaviour of $S''$ at the critical points $\omega^+$ and $\omega^-$ is evident from the above analysis. The fixed point $\omega^+$ represents the earlier epoch and
$\omega^-$ represents the later epoch for $s=1/2.$ However, as a matter of simple academic interest,
the evolution equation of $S''$ at the critical points can be expressed as,
\begin{equation}
\label{eqn:S''_P+}
S''_{\omega^\pm}=\frac{2\pi c^2}{l_P^2 } (1+2q^\pm )
(1+q^\pm)^4 a^{2q^\pm}.
\end{equation}
\begin{figure}
	\centering
	\includegraphics{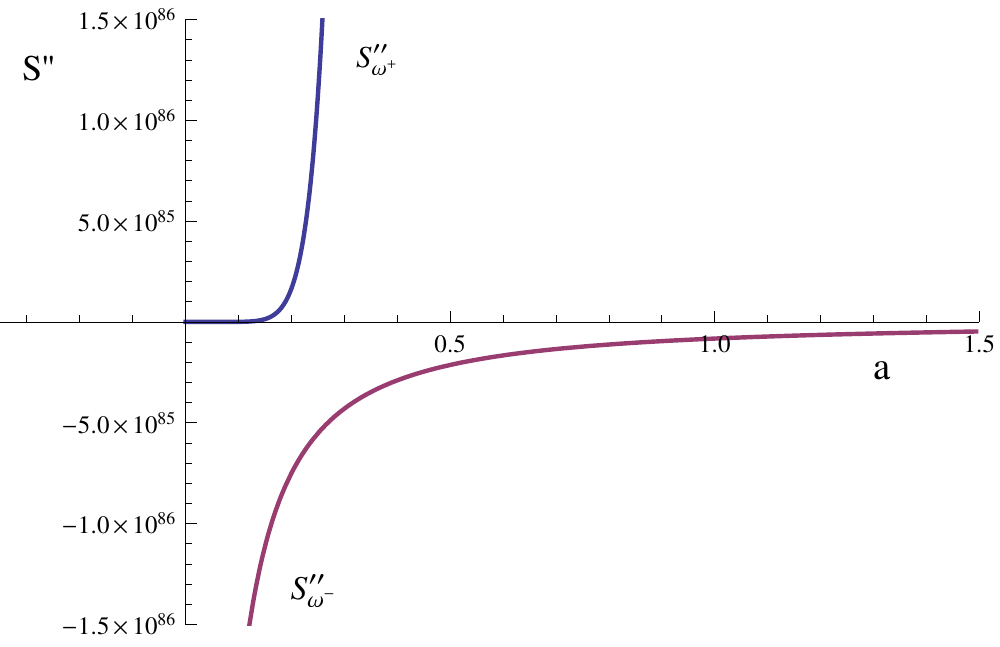}
	\caption{The evolution of $S'' $ in units of $k_B$ with scale factor $a$ at the critical points $\omega^\pm$ for the best estimated value of the model parameters, when $s=1/2$}
	\label{plot:S''_omega+-}
\end{figure}
From the
figure \ref{plot:S''_omega+-}, it is clear that convexity condition is violated at the critical point $\omega^+$ but satisfied at
$\omega^-$ as expected.
This indicates that the first critical point $\omega^+$ is an unstable thermodynamic equilibrium and the second point $\omega^-$
is thermodynamically stable.

Fro $s>1/2$ the second derivative of entropy at the critical points $P_1$ can be obtained using (\ref{eqn:HP1P2}) and (\ref{eqn:S''_general}) as,
\begin{equation}
\label{eqn:S''_P1P2}
\frac{d^2 S}{d{\tilde{\tau}}^2}=\frac{4\pi c^2}{l_p^2}(1+q)^3 e^{2(1+q)\tau}.
\end{equation}
For $P_1,$ which represents the
deceleration parameter $q>0,$ the term $\frac{d^2 S_{P_2}}{d{\tilde{\tau}}^2}>0,$ and the convexity condition is hence violated.
At the equilibrium point $P_2,$ representing a de Sitter epoch, we observe that the $S''$ will vanish hence the convexity
condition is not strictly satisfied.
These results are summarised in table \ref{Table:FISThermalproperties}.

\begin{table}	
	\center{\begin{tabular}{@{}lcccc@{}}\mbox{} \\\hline\hline\\                                 
			Critical points $\downarrow$ &  LSL & GSL & $S''<0$ \\\hline	
			$\omega^+$ & Yes & Yes & No  \\
			$\omega^-$ & Yes & Yes & Yes \\
			$P_1$  & Yes  & Yes  & No \\
			$P_2$ & Yes & Yes & No  \\\hline\hline
	\end{tabular}}\\
	\caption{Thermal properties of the critical points $\omega^+, \omega^-,  P_1$ and $P_2$ of the model. In the table LSL and GSL denote the local second law and generalized second law of  thermodynamics respectively and $S''<0$ is the convexity condition of entropy.} 
	\label{Table:FISThermalproperties}		\end{table}

A causal dissipative model for $s=1/2$ having barotropic equation of state for matter, $p=\omega \rho,$ with $\omega$ varying in the range $0<\omega<1,$
has been analyses in reference \cite{Cruz1}. The authors assumed an ansatz for
Hubble parameter of the form, $H(t>t_s)=|A|/(t-t_s),$ where $|A|$ is a positive coefficient depending on $\omega$ and the viscous coefficient
$\xi.$ The IS transport equation will then give rise to a quadratic equation for $|A|,$ with two possible solutions, say $|A|_+$
and $|A|_-.$ By considering only the solution corresponding to $|A|_+,$
the authors have argued that
the GSL is satisfied in both the prior decelerated and later accelerated phases, but the convexity condition is satisfied by the early phase but violated
in the later accelerated epoch.
In contrast to this, the analytical solutions that we have obtained
for the
IS equation with $s=1/2$ and zero barotropic pressure, i.e. $\omega=0,$
predicts
an early decelerated epoch which satisfies GSL but violate convexity condition and a late accelerated phase which satisfies both GSL and convexity conditions.
\section{Conclusions}
\label{sec:4}
In this work we have analysed the dynamical system behaviour and thermodynamic characteristics of the late universe with a dissipative
fluid using the full Israel-Stewart theory.
Assuming the bulk viscosity as $\xi = \alpha \rho^s,$
we consider two separate cases one with $s=1/2$ and the other with $s\neq 1/2.$

For $s=1/2$ we could obtain an analytical solution for the Hubble parameter, by which the model implies a prior decelerated epoch and a late accelerated epoch.
The corresponding phase space is found to be reduced to a one dimensional one with
two fixed points $\omega^+$ and $\omega^-$ corresponding to an early decelerated and late accelerating phases respectively.
It emerges from our analysis that $\omega^+$ is a past attractor
hence unstable while the late accelerating epoch corresponding to $\omega^-$ is a stable one. We have seen that
the effective equation of state indicates a stiff nature for the viscous matter in the neighbourhood of the fixed point $\omega^+.$ At $\omega^-,$ corresponding to the
the late accelerating phase
the equation of state become \textbf{$\omega \sim -0.79,$} implying a quintessence nature but not pure de Sitter. Regarding energy conditions, it is easy to see that
both fixed points satisfy the dominant energy condition, but
the strong energy condition is satisfied only by $\omega^+$ as a consequence of its decelerating nature. When $s=1/2,$ a general behaviour of the bulk viscous model has analyzed using a general relaxation time expression (\ref{eqn:relaxationtimegeneral}), by varying $\epsilon$ and barotropic index $\gamma.$ For the best estimated parameter values, the models exhibits the quintessence evolution, however the late phase stabilizing values of equation state is close to $-1.$ 

The next choice is $s\neq1/2.$ When $s<1/2$ there exist two critical points in which the first one is corresponding to static universe, while the second is
giving a de Sitter epoch. Since the first static solution prohibits any further evolution, the case fails to explain the conventional evolution of the universe. As a result
the case $s<1/2$ is not worth studying and can be ruled out. For the case $s>1/2$ the phase space becomes two dimensional with coordinates $h=H^{1-2s}$ and $\omega$ and having
two critical points, out of which the first one, $P_1$ represents a decelerated epoch and the second one $P_2$ indicating the de Sitter epoch.
Our analysis on stability
shows that, $P_1$ is a saddle point and $P_2$ is a repeller, hence unstable. Hence a stable evolution towards an end de Sitter epoch is unlikely for $s>1/2.$
In the energy condition analysis, for
the case $s> 1/2,$  we found that
both SEC and DEC are satisfied at $P_1,$ which is corresponding to a prior decelerated phase of expansion. In the case of $P_2,$ DEC is satisfied while SEC is violated as
is representing a late accelerated epoch.

In the analysis of the thermodynamic characteristics, we have shown that, for $s=1/2$ the model satisfies the GSL, $S^{\prime}\geq 0$ through out the evolution
and obeys the convexity condition $S^{\prime\prime}<0$
in the long run of the expansion. Then as matter of fact we verified, in the case of the corresponding fixed points, that GSL is valid at both the critical
points $\omega^+$ and $\omega^-$ but convexity condition is satisfied only by the later critical point $\omega^-.$
This indicates that the expansion is tending towards a state of a maximum
entropy as in the evolution of an ordinary macroscopic system.

For $s\neq1/2$
we restrict the thermodynamic analysis to the case $s>1/2.$
The GSL is valid at both the critical points in this case.
Among these we already noted that $P_1$ represents
a prior decelerated epoch and $P_2$ corresponds the future
de Sitter epoch. Regarding the convexity condition, our result is that, it is violated at both the fixed points
prohibiting an upper bound for the
growth of entropy. Hence the case $s>1/2$ does not imply a stable thermodynamic evolution. This is in line with
the dynamical system behaviour of this case also, by which we found both critical points are unstable.

To summarise, for the choice $s=1/2,$ the present dissipative model described using the Israel-Stewart theory predicts a stable evolution of the late universe with
prior decelerated epoch followed by an accelerated epoch. The GSL is valid throughout the evolution and the entropy is bounded
for the end phase with the convexity condition satisfied. We can also infer that the thermal properties of the bulk viscous universe, especially the entropy, exhibits drastic change 
during the phase transition period. For the choice $s\neq1/2,$ the case with $s<1/2$ can be ruled out since it predicts an evolution not in conformity with the conventional evolution of the universe.
The case $s>1/2$ predicts a prior decelerated and a late accelerated phase, but fails to predict a stable evolution.  Finally we would like to comment that apart from explaining the late acceleration, the viscous models are found successful in certain other areas too. For example 
in reference \cite{anand1} the authors have shown 
that, a  very small viscosity of the order of $ 10^{-6}$Pa sec (1$\sigma$ level)
in the dark matter sector can cure the $\sigma_8-\Omega_m$ tension ($\sigma_8$ is the r.m.s. fluctuations of perturbations at $8h^{-1}$Mpc scale) and the $H_0 -\Omega_m$ tension occurred when one analyse the Planck CMB parameters using the standard $\Lambda$CDM model.

\section*{Acknowledgments}

We are also thankful to IUCAA, Pune for the hospitality during the visits. We are also thankful to the referees for the comments, which helped to improve the manuscript.Authors are grateful to Prof. M. Sabir for the careful reading of the manuscript. 
Author JMND acknowledges UGC - BSR for the fellowship, author KPB acknowledges KSCSTE, Government of Kerala for financial assistance and author AS is 
thankful to DST for fellowship through the INSPIRE fellowship. \\[0.05in]

\noindent{\bf Appendix}\\
{ \textbf{The possibilities of attaining a pure de Sitter epoch for $s=1/2$}

	So far our analysis have shown that the model allows an asymptotic value for equation of state around, $\omega = -0.79.$ Now we check possibilities 
	of improving this value so that the model can predict a pure de Sitter epoch with $\omega=-1$ in the long run.
	Previously we took the relaxation time as $\tau = \alpha \rho^{s-1}$ with $s=1/2.$ Since this doesn't gives an asymptotic de Sitter epoch, let us 
	relax this condition by assuming a more general relation for the relaxation time as \cite{Maartens2}, 
	\begin{equation}
	\label{eqn:relaxationtimegeneral}
	\tau = \frac{\alpha}{\epsilon \gamma (2 - \gamma)} \rho^{s-1}. 
	\end{equation}
	We first fix the barotropic index as, $\gamma=1$ and allow to vary the parameter $\epsilon$ in the range  
	$0 < \epsilon \leq 1.$ \\
	The solution for the Hubble parameter is obtained in \cite{Cruz2}, which have the same form as in the previous case,
	\begin{equation}
	H=H_0(C_1 a^{-m_1}+C_2a^{-m_2}),
	\end{equation}
	but with different coefficients,
	\begin{equation}
	C_{1;2}=\frac{\pm \epsilon+\sqrt{6\epsilon\alpha^2+\epsilon^2}\mp\sqrt{3}\alpha\tilde{\Pi_0}}{2\sqrt{6\epsilon\alpha^2+\epsilon^2}},
	\end{equation}
	\begin{equation}
	m_{1;2}=\frac{\sqrt{3}}{2\alpha}\left(\sqrt{3}\alpha+\epsilon\mp\sqrt{6\epsilon\alpha^2+\epsilon^2}\right),
	\end{equation} 
	which are satisfying the conditions $C_1+C_2=1$ and $m_2>m_1.$  Using the Supernovae type Ia data we have extracted the parameter values 
	in the present case as $\alpha=169.50,$ $\tilde{\Pi_0}=-0.70,$ $\epsilon=0.39$ and $H_0=69.99$ with $\chi^2_{d.o.f.}=0.97.$  
	We have then concentrated on the evolution of the equation of state which can be analytically obtained as,
	\begin{equation}
	\label{eqn:eqnofstateFISwithe}
	\omega=-1+\frac{2(C_1 m_1 a^{-m_1}+C_2 m_2 a^{-m_2})}{3(C_1 a^{-m_1}+C_2a^{-m_2})}.
	\end{equation}
	To get the late phase behaviour, consider the asymptotic limit of equation of state parameter (\ref{eqn:eqnofstateFISwithe}) when the scale factor $a\rightarrow\infty.$ In the late phase evolution, the equation of state parameter (\ref{eqn:eqnofstateFISwithe})
	can takes the form,
	\begin{equation}
	\omega\sim -1+\frac{2}{3}m_1.
	\end{equation}
	For the new best estimated parameter values, the constant, $m_1=0.18,$ and hence the equation of state parameter will stabilizes around $\omega\sim -0.88.$
	So the values has been improved slightly but still not represent a pure de Sitter case.
	
	As a further move we extend the analysis by varying the parameter $\gamma$ also. By considering this, a more general solution 
	for the Hubble parameter can be obtained as discussed in \cite{Cruz3} as,
	\begin{equation}
	\label{eqn:Hubbleparameterwithgamma }
	H=C_3 (1+z)^{\alpha'} cosh^\gamma\left[\beta(ln(1+z)+C_4)\right],
	\end{equation}
	where \begin{equation}
	\nonumber
	C_3=H_0\left[1-\frac{(q_0+1-\alpha')^2}{\gamma^2 \beta^2}\right]^{\gamma/2},
	\end{equation}
	\begin{equation}
	\nonumber
	C_4=\frac{1}{\beta}arctanh\left[\frac{(q_0+1)-\alpha'}{\gamma\beta}\right],
	\end{equation}
	\begin{equation}
	\nonumber
	\alpha'=\frac{\sqrt{3}\gamma}{2 \xi_0}\left[\sqrt{3}\xi_0+\epsilon\gamma(2-\gamma)\right],
	\end{equation}
	\begin{equation}
	\nonumber
	\beta=\frac{\sqrt{3}}{2\xi_0}\sqrt{6\xi_0^2\epsilon(2-\gamma)+\epsilon^2\gamma^2(2-\gamma)^2}.
	\end{equation}
	where $q_0$ is the present value of deceleration parameter and $\xi_0$ is the viscosity constant parameter ($\xi_0=\alpha$ in our analysis). Following reference \cite{Cruz3}, 
	the model parameters take the values as $\xi_0=245.2,$ $\epsilon=0.601$ and $\gamma=1.26$ with $\chi^2_{d.o.f.}=1.07$ for $H_0=70km/Mpcs$ and $q_0=-0.60.$ 
	Using the equation of parameter evaluating equation \cite{EPJC1}, we have obtained the 
	the asymptotic limit of equation of state parameter for the estimated parameter values, when $a\rightarrow \infty,$ the equation of state 
	$\omega\sim -0.93,$ and is very close to the de Sitter epoch value.
	Therefore, we have conclude that, even though the model will not attain the pure de Sitter epoch ($\omega=-1$) as the end phase,  
	it attains a quintessence 
	epoch which very close to the de Sitter phase. \\[0.2in]}


\begin{thebibliography}{}
	\bibitem{Riess} A. G. Riess et al. (Supernova Search Team Collaboration)  1998 { \it Astron. J.} { \bf  116} 1009
	\bibitem{Perlmutter} S. Perlmutter et al. (Supernova Cosmology Project Collaboration) 1999  { \it Astrophys. J.} { \bf  517} 565
	\bibitem{Bennett} C. L. Bennett et. al. (WMAP Collaboration) 2003
	{ \it Astrophys. J. Suppl.} { \bf  148} 1
	\bibitem{Riess2} A. G. Riess et. al. (Supernova Search Team Collaboration) 2004
	{ \it Astrophys. J.} { \bf  607} 665
	\bibitem{Tegmark} M. Tegmark et. al. (SDSS Collaboration) 2004
	{ \it Phys. Rev. D} { \bf  69} 103501  
	\bibitem{Seljak} U. Seljak et. al. (SDSS Collaboration) 2005
	{ \it Phys. Rev. D} { \bf  71} 103515
	\bibitem{Komatsu} E. Komatsu et. al. (WMAP Collaboration) 2011
	{ \it Astrophys. J. Suppl.} { \bf  192} 18 
	\bibitem{Sami1} E. J. Copeland, M. Sami and S. Tsujikawa, 2006 { \it Int. J. Mod. Phys. D} { \bf  15} 1753
	\bibitem{Wang} L. Wang, R. R. Caldwell, J. P. Ostriker and P. J. Steinhardt, 2000 { \it Astrophys. J.} { \bf  530}  17
	\bibitem{Caldwell} R. R. Caldwell, 2002 { \it Phys. Lett. B} { \bf  545} 23
	\bibitem{Bamba} K. Bamba, K. Capozziello, S. Nojiri and S. D. Odintsov, 2012 { \it Astrophys. Space Sci.} { \bf  342} 155 
	\bibitem{wang1} B. Wang, E. Abdalla, F. Atrio-Barandela and D. Pavon, 2016 { \it Rep. Prog. Phys.} { \bf  79} 096901
	\bibitem{Dvali} G. R. Dvali, G. Gabadadze and M. Porrati, 2000 { \it Phys. Lett. B} { \bf  484}  112
	\bibitem{Freese} K. Freese and M. Lewis, 2002 { \it Phys. Lett. B} { \bf  540} 1
	\bibitem{Brevik1} I. Brevik and O. Gorbunova, 2005 { \it Gen. Rel. Grav.} { \bf  37} 2039  
	\bibitem{Brevik2} I. Brevik, O. Gorbunova and Y. A. Shaido, 2005 { \it Int. J. Mod. Phys. D} { \bf  14} 1899
	\bibitem{Avelino1} A. Avelino and U. Nucamendi, 2009 { \it JCAP} { \bf  04} 006 
	\bibitem{Athira} Athira Sasidharan and Titus K. Mathew, 2015 { \it Eur. Phys. J. C.} { \bf  75} 348 
	\bibitem{Misner} C. W. Misner, K. S. Thorne and J. A. Wheeler, 1973 {\it Gravitation} ( U.S.A: W. H. Freeman and Company) 
	\bibitem{Weinberg} S. Weinberg, 1972 {\it{Gravitation and Cosmology: Principles and Applications of the General Theory of Relativity}} (New York: Wiley) 
	\bibitem{Weinberg1} S. Weinberg, 1989 { \it Rev. Mod. Phys.} { \bf  61} 1 
	\bibitem{anand1} S. Anand, P. Chaubal, A. Mazundar and S. Mohanty, 2017 { \it J. Cosmol. Astropart. Phys.,} { \bf  11}  005
	\bibitem{Wilson} J. R. Wilson, G. J. Mathews and G. M. Fuller, 2007 { \it Phys. Rev. D} { \bf  75}  043521
	\bibitem{Okumura} H. Okumura and F. Yonezawa, 2003 { \it Physica A} { \bf  321} 207 
	\bibitem{Mathews} G. J. Mathews, N. Q. Lan and C. Kolda, 2008
	{ \it Phys. Rev. D} { \bf  78} 043525 
	\bibitem{Avelino2} A. Avelino and U. Nucamendi, 2010 { \it JCAP} { \bf  08}  009 
	
	\bibitem{Weinberg2} S. Weinberg, 1971 { \it Astophys. J} { \bf  168} 175 
	\bibitem{Schweizer} M. A. Schweizer, 1982 { \it Astrophys. J} { \bf  258} 798 
	\bibitem{Udey} N. Udey and W. Israel, 1982 { \it Mon. Not. R. Astron. Soc.} { \bf  199} 1137
	\bibitem{Zimdahl3} W. Zimdahl, 1996 { \it Mon. Not. R. Astron. Soc.} { \bf  280} 1239 
	\bibitem{Murphy} G.L. Murphy, 1973 { \it Phys. Rev. D} { \bf  8} 4231 
	\bibitem{Turok} N. Turok, 1988 { \it Phys. Rev. Lett.} { \bf  60} 549 
	\bibitem{Zimdahl4} W. Zimdahl and D. Pavon, 1993 { \it Phys. Lett. A} { \bf  175} 57 
	\bibitem{Eckart} C. Eckart, 1940 { \it Phys. Rev.} { \bf  58} 919
	\bibitem{Coley2} A. A. Coley and R. J. van den Hoogen, 1995 { \it Class. Quantum Grav.} { \bf  12}  1977
	\bibitem{Israel1} W. Israel, 1976 { \it Ann. Phys. (N. Y.)} { \bf  100} 310
	\bibitem{Hiscock1} W. A. Hiscock and L. Lindblom, 1985 { \it Phys. Rev. D} { \bf  31} 725
	\bibitem{Fabris} J. C. Fabris, S. V. B. Goncalves and R. de Sa Ribeiro, 2006 { \it Gen. Relativ. Gravit.} { \bf  38} 495
	\bibitem{Barrow1} J. D. Barrow, 1986 { \it Phys. Lett. B} { \bf  180} 335 
	\bibitem{Colistete} R. Colistete, J. C. Fabris, J. Tossa and W. Zimdahl, 2007 { \it Phys. Rev. D} { \bf  76} 103516
	\bibitem{Athira2} Athira Sasidharan and Titus K. Mathew,  2016 { \it JHEP} { \bf  06} 138 
	\bibitem{IsraelStewart} W. Israel and J. M. Stewart, 1979 { \it Annals Phys.} { \bf  118} 341
	\bibitem{IsraelStewart2} W. Israel and J. M. Stewart, 1979 { \it Proc. Roy. Soc. Lond. A} { \bf  365} 43 
	\bibitem{Piattella} O. F. Piattella, J. C. Fabris and W. Zimdahl, 2011 { \it JCAP,} { \bf  05} 029
	\bibitem{Padmanabhan} T. Padmanabhan and S. M. Chitre, 1987 { \it Phys. Lett. A} { \bf  120} 443
	\bibitem{Prisco} A. Di Prisco, L. Herrera, and Ib\'a\~nez, J.,  2000 { \it Phys. Rev. D} { \bf  63}  023501
	\bibitem{EPJC1} Jerin Mohan N D, Athira Sasidharan and Titus K. Mathew, 2017 { \it Euro. Phys. J. C.} { \bf  77} 849
	\bibitem{Brevikrev1} I. Brevik, O. Green, J. de Haro, S. D. Odintsov and E. N. Saridakis, 2017 { \it Int. Nat. J. Mod. Phys. D} { \bf  26} 1730024
	\bibitem{Pavon1} Diego Pavon and Ninfa Radicella, 2013 { \it Gen. Relativ. Gravit} { \bf  45}  63 
	\bibitem{Cruz1} M. Cruz, N. Cruz and S. Lepe, 2017 { \it Phys. Rev. D} { \bf  96} 124020 
	\bibitem{Maartens} R. Maartens, 1995 { \it Class. Quantum Grav.} { \bf  12}  1455
	\bibitem{ChimentoJacubi} L. P. Chimento and A. S. Jacubi, 1997 { \it Class. Quantum Grav.} { \bf  14} 1811
	\bibitem{Ellis} Wainwright J and Ellis G F R, 1997 {\it{Dynamical Systems in Cosmology}} (Cambridge: Cambridge
	University Press)
	\bibitem{U.Alam} U. Alam, V. Sahini, A. A. Starobinsky, 2004 { \it JCAP},{ \bf  0406} 008 
	\bibitem{Awad1} A. Awad, W. E. Hanafy, G. Nashed, and E. N. Saridakis, 2018 { \it JCAP} { \bf  2018} 052
	\bibitem{Visser} M. Visser, 1997 { \it Science} { \bf  276} 88
	\bibitem{Pavon12} D. Pavon, 1990  { \it Classical and Quantum Gravity} { \bf  7} 487 
	\bibitem{Barrow2} J. D. Barrow, 1987 { \it Phys. Lett. B} { \bf  183} 285 
	\bibitem{Krishna} Krishna P. B. and Titus K. Mathew, 2017 { \it Phys. Rev. D} { \bf  96} 063513
	\bibitem{Davis} P. C. W. Davis, 1987 { \it Class. Quantum Gravity} { \bf  4}, L225 
	\bibitem{Callen1} H B Callen, 1985 {\it Thermodynamics and an Introduction to Thermostatistics} (New York: John Wiley)
	\bibitem{Maartens2} R. Maartens, {\it arXiv:astro-ph/9609119}
	\bibitem{Cruz2} N. Cruz, E. Gonzalez, G. Palma, 2018 {\it arXive:1812.05009v3}
	\bibitem{Cruz3} N. Cruz, E. Gonzalez, G. Palma, 2019 {\it arXive:1906.04570}\\ \\
\end{thebibliography}
\end{document}